\documentclass[twocolumn,prd,unsortedaddress,superscriptaddress,showpacs,a4paper,nofootinbib]{revtex4-1}
\usepackage[T1]{fontenc}
\usepackage{graphicx}
\usepackage{indentfirst}
\usepackage{pdfpages}
\usepackage{multirow}
\usepackage{amssymb}
\usepackage{amsfonts}
\usepackage{amsmath}
\usepackage{epsfig}

\def\beq{\begin{equation}}
\def\enq{\end{equation}}
\def\beqa{\begin{eqnarray}}
\def\enqa{\end{eqnarray}}

\def\MeV{\nobreak\,\mbox{MeV}}
\def\GeV{\nobreak\,\mbox{GeV}}

\def\Tr{\mbox{Tr}}

\def\qq{\lag\bar{q}q\rag}

\def\qGq{\lag\bar{q}Gq\rag}
\def\GG{\lag G^2 \rag}
\def\gGG{\lag g_s^2 G^2 \rag}
\def\GGG{\lag G^3\rag}
\def\gGGG{\lag g_s^3 G^3 \rag}
\def\qqqq{\lag\bar{q}q\bar{q}q\rag}
\def\pli{p^\prime}

\def\la{\lambda}

\def\Ga{\Gamma}

\def\de{\delta}
\def\al{\alpha}

\def\lb{\label}
\def\nn{\nonumber}

\newcommand{\rag}{\rangle}
\newcommand{\lag}{\langle}

\newcommand{\rsup}[1]{\mbox{\tiny $#1$}}

\newcommand{\degree}{\ensuremath{^\circ}}

\begin{document}
\title{$Y(3940)$ as a Mixed Charmonium-Molecule State}
\author{R.M. Albuquerque}
\email{raphael@ift.unesp.br}
\affiliation{Institute  for  Theoretical  Physics, S\~ao  Paulo  State
  University  (IFT-UNESP)\\  R.\,Dr.\,Bento\,Teobaldo\,Ferraz,  271  -
  Bl.\,II Sala 207, 01140-070 S\~ao Paulo/SP - Brasil}
\author{J.M. Dias}
\email{jdias@if.usp.br}
\author{M. Nielsen}
\email{mnielsen@if.usp.br}
\affiliation{Instituto de F\'{\i}sica, Universidade de S\~{a}o Paulo, 
C.P. 66318, 05389-970 S\~{a}o Paulo, SP, Brasil}
\author{C.M. Zanetti}
\email{carina.zanetti@gmail.com}
\affiliation{ Faculdade de Tecnologia, Universidade do Estado do Rio de 
Janeiro, Rod. Presidente Dutra Km 298, P\'olo Industrial, 27537-000 , 
Resende, RJ, Brasil}

\begin{abstract}
Using the QCD sum rules approach we  study the mass and decay width of
the channel $J/\psi\:\omega$ for the  $Y(3940)$ state, assuming that it
can  be   described  by  a  mixed   charmonium-molecule  scalar  state
$(\chi_{c0})\!-\!(D^\ast \!\bar{D}^\ast)$ current with $J^{PC}=0^{++}$
quantum   numbers.   For   the   mixing  angle   $\theta  =   (76.0\pm
5.0)\degree$,  we  obtain  the  value  $M_{Y}  =  (3.95  \pm  0.11)  ~
\mbox{GeV}$  for  the  mass,  which  is in  good  agreement  with  the
experimental mass  of the $Y(3940)$  state.  For the decay  width into
the  channel $Y  \to J/\psi\:\omega$  we  find the  value $\Ga_{Y  \to
  J/\psi\:\omega} = (1.7\pm 0.6) \MeV$, which is also compatible
with  the  experimental  data.   We thus  conclude  that  the  present
description of the $Y(3940)$ as a mixed charmonium-molecule state is a
possible scenario to explain the structure of this state.
\end{abstract}

\pacs{11.55.Hx, 12.38.Lg , 12.39.-x}
\maketitle

\section{Introduction}
In the past years  several states in the region of  mass of about 3940
MeV has been observed in  different processes of production and decay.
The state  $X(3915)$ was  observed by {\sc Belle} Collaboration  in the
process $\gamma\gamma\to\,J/\psi\,\omega$ \cite{belleX3915}, with a mass
$m=(3915 \pm 3 \pm 2)$ MeV and  total width $\Gamma=(17 \pm 10 \pm 3)$
MeV.  In  addition, the  observation of the state $Y(3940)$  has been
made by  {\sc Belle} Collaboration in  the decay $B\to  (J/\psi\,\omega)K $,
with a mass $m=3943\pm11({\mbox{{\it stat}}})\pm13({\mbox{{\it syst}}})$
MeV and decay width $\Gamma=87\pm22({\mbox{\it stat}})\pm26({\mbox{\it
    syst}})$ MeV \cite{belleY3940}.  Afterwards, this state has been also
observed  in the process $B\to  (J/\psi\,\omega)K$  by  {\sc Babar}
Collaboration,     with     a     slightly     smaller     mass     of
$m=3914.6^{+3.9}_{-3.4}~({\mbox{{\it       stat}}})~\pm~2.0({\mbox{\it
    syst}})$   MeV    and   width   $\Gamma=34^{+12}_{-8}~({\mbox{{\it
      stat}}})~\pm~5.0({\mbox{\it  syst}})$ MeV  \cite{babarY3940}. In
the  same mass  region,  the  state $Z(3930)$  was  discovered in  the
process $\gamma\gamma\to\,D\bar{D}$,  that is generally linked  to the
charmonium state $\chi_{c2}(2P)$ \cite{belleZ3940,babarZ3940}.

The  proximity of  the  masses could indicate that  all  these states  are
connected to the same particle observed in different processes.  There
are  evidences, however,  that   the  two reported  states,  $Y(3940)$  and
$X(3915)$, could be interpreted as molecular states.  The $X(3915)$ state
has a larger product of the two-photon width times the decay branching
fraction than usually expected for  charmonium states, as noted in
Ref. \cite{babar2}.  Regarding the $Y(3940)$,  the lower limit for the
decay channel  $J/\psi\:\omega$ has been  estimated to be  $\Gamma> 1$
MeV,  which  is  large  for  a channel  that  is  OZI  suppressed  for
conventional charmonium  states \cite{godfrey,godfrey2}.   These facts
suggests that  these states  cannot be  interpreted as  a conventional
$c\bar{c}$  state.   In  Ref.~\cite{liu},  it  was  proposed  that  the
$Y(3940)$ can be a molecular state $D^*\bar{D}^*$, with quantum numbers
$J^{PC}=0^{++}$ or $2^{++}$.  It was also concluded that the $Y(3940)$
must   be  the   molecular   partner  of   the   state  $Y(4140)$,   a
$D^*_s\bar{D}^*_s$ molecule.   This interpretation has been  tested in
several    approaches,    such   as    phenomenological    lagrangians
\cite{gutsche}   and    vector-meson   dominance    \cite{oset}.    In
Ref.~\cite{albuquerque},  the $Y$ state was  studied  with  QCD Sum  Rules
(QCDSR) method \cite{svz,rry,SNB} as  a $D^*\bar{D}^*$ molecule with  quantum numbers $0^{++}$
and  the  mass   obtained  was  $m_{D^*\bar{D}^*}=(4.13\pm0.10)$  MeV,
failing to reproduce the experimental mass of the state.

In the present work we revisit the study  of the $Y(3940)$ 
within QCDSR approach, using a mixed charmonium-molecule current.
The  prescription of a mixture of two- and four-quarks states has been 
successfully implemented for other states in the framework of sum rules. 
Following the work of Ref.\,\cite{oka} that was applied in the light quark sector, 
the authors in Refs.~\cite{matheusX3872,x3872rad,x3872prod} described the 
$X(3872)$ state as a  molecule-charmonium state, implementing the mixing of 
the current and extending it to the charm sector.  In  these  works  the
mass and decay width for the channels $J/\psi+(2\pi,3\pi,\gamma)$ and the
production in  $B$-meson decays were estimated in a good agreement with
the experimental values. Another state that  was studied as a mixture was
the $Y(4260)$.   In Ref.~\cite{diasY4260}, the $Y(4260)$  was described
as  a  tetraquark-charmonium mixed state,  and the mass and decay width
estimated are also consistent with the experimental values.

In the following  sections we use the QCDSR approach to describe the 
$Y(3940)$ as   a   mixing   between the $\chi_{c0}$  charmonium  and 
the $D^\ast\!\bar{D}^\ast$ molecule, with $J^{PC}=0^{++}$.  
We  obtain the mass for this state and the decay width in the 
channel $Y\to J/\psi \:\omega$.

\section{Mixed Hadronic Current}
\label{mixoperator}

In order to evaluate  the sum rule for the $Y(3940)$  state as a mixed
$(\chi_{c0})\!-\!(D^\ast  \!\bar{D}^\ast)$   state,  with   $J^{PC}  =
0^{++}$,  one employs  the following  hadronic current  
\beqa j  &=& a
\:\mbox{cos}\:\theta   ~j_{_{\chi_{c0}}}   ~+~   \mbox{sin}   \:\theta
~j_{_{D^\ast \!D^\ast}}
  \label{mixj}
\enqa  
where $\theta$  is an  arbitrary  mixing angle.  The meson  and
molecule currents are, respectively,  given by: 
\beqa j_{_{\chi_{c0}}}
&=&  \bar{c}_k  c_k  \\  j_{_{D^\ast  \!D^\ast}}  &=&  \left(\bar{q}_i
\gamma_\mu c_i \right) \left(\bar{c}_j \gamma^\mu q_j \right)~~.
  \label{curr}
\enqa
Notice   that  the   normalization   factor  $a$   is  introduced   in
Eq.~(\ref{mixj}) for ensuring that the  mixed current can be evaluated
at the same Fock space. Usually, one sets 
\cite{oka,matheusX3872,x3872rad,x3872prod}
\beq
  a =  - \frac{\qq}{\sqrt{2}} ~.
\enq
Then,  evaluating  the  two-  and  three-point  correlation  functions
altogether with Eq.~(\ref{mixj})  one can estimate the  mass and decay
width of the mixed $(\chi_{c0})\!-\!(D^\ast \!\bar{D}^\ast)$ state.

\section{Two-Point Correlation Function}

To obtain the mass of a hadronic state using the QCDSR approach, one has
to evaluate the two-point correlation function
\begin{eqnarray}
  \Pi(q) &=& i \int d^4x ~e^{i q\cdot x} \langle 0| \:T[ j(x) j^\dagger (0) ] \:|0 \rangle
  \label{fc}
\end{eqnarray}

According to the quark-hadron duality principle, Eq.~(\ref{fc}) can be evaluated
in two ways: the phenomenological side and the QCD side.
The  phenomenological side is calculated by inserting,
in Eq.~(\ref{fc}), a  complete set of intermediate  states, $Y$, which
 couple   to   the   hadronic   current   in   Eq.~(\ref{mixj}).
Parametrizing this  coupling through a generic  parameter $\lambda_Y$,
one defines
\begin{equation}
\langle 0| \:j\: | Y \rangle =  \lambda_Y ~.
\label{coupling}
\end{equation}
Using Eq.~(\ref{coupling}) and after  some algebraic manipulation, one
can write the phenomenological side of Eq.~(\ref{fc}) as
\begin{equation}
  \Pi^{\rsup{PHEN}}(q) = \frac{\lambda_Y^{2}}{M_{Y}^{2} - q^{2}} + 
  \int\limits_0^\infty \!ds \:\frac{\rho^{\rsup{cont}}(s)}{s-q^2}
\label{fcphen}
\end{equation}
where  $M_Y$ is  the  mixed $(\chi_{c0})\!-\!(D^\ast  \!\bar{D}^\ast)$
ground state mass and the second term in the RHS of Eq.~(\ref{fcphen})
denotes the continuum (or higher resonance) contributions. As usual in
a QCDSR approach, it is assumed that the continuum contribution to the
spectral   density,  $\rho^{\rsup{cont}}(s)$   in  Eq.~(\ref{fcphen}),
vanishes below a certain threshold  $s_0$. Above this threshold, it is
assumed that  the result coincides  with the  one obtained in  the OPE
side. Therefore, one uses the ansatz \cite{io1}
\begin{eqnarray}
  \rho^{\rsup{cont}}(s) &=& \rho^{\rsup{OPE}}(s) \:\Theta(s - s_0)
\end{eqnarray}
where $\Theta(s - s_0)$ is the Heaviside step function.

In  the OPE  side, one calculates  the correlation
function  in  terms of  quark  and  gluon  fields using  the  Wilson's
operator  product expansion  (OPE).  This is also called the OPE side.
Then, inserting  Eq.~(\ref{mixj})
into the above equation, one obtains
\begin{eqnarray}
\Pi^{\rsup{OPE}}(q) &=& i \!\int \!\!d^4x \,e^{i q\cdot x} \Bigg\{ \!\frac{1}{2} 
  \qq^2 \mbox{cos}^2\theta ~\Pi_{_{\chi_{c0}}} + 
  \mbox{sin}^2\theta ~\Pi_{_{D^\ast \! D^\ast}} \nn\\
  && -~ \frac{\qq}{\sqrt{2}} \:\mbox{sin}\,\theta \:\mbox{cos}\,\theta ~
  \Big[ \Pi_{_{mix}} + \Pi_{_{mix}}^\ast \Big] \Bigg\} 
  \label{fcope}
\end{eqnarray}
where the  $\Pi_{_{\chi_{c0}}}(x)$ and $\Pi_{_{D^\ast  \! D^\ast}}(x)$
functions  are,   respectively,  the  correlation  functions   of  the
$\chi_{c0}$ meson  and the $D^\ast D^\ast  ~(0^{++})$ molecular state,
which have  been calculated in  other works \cite{albuquerque, rry}.  Thus, one
only has to calculate the $\Pi_{_{mix}}(x)$ and $\Pi_{_{mix}}^\ast(x)$
functions defined as follows:
\beqa
  \Pi_{_{mix}}(x) &=&  \langle 0| \:T[ j_{_{\chi_{c0}}}(x) j_{_{D^\ast \!D^\ast}}^\dagger (0) ] \:|0 \rangle \nn\\
  &=& -\Tr \bigg[ S^q_{ji}(0) \:\gamma_\mu \: S^{c}_{ik}(-x) S^c_{kj}(x) \:\gamma^\mu \bigg] \\ &&\nn\\
  \Pi_{_{mix}}^\ast(x) &=&  \langle 0| \:T[ j_{_{D^\ast \!D^\ast}}(x) j_{_{\chi_{c0}}}^\dagger (0) ] \:|0 \rangle \nn\\
  &=& -\Tr \bigg[ S^q_{ji}(0) \:\gamma_\mu \: S^{c}_{ik}(x) S^c_{kj}(-x) \:\gamma^\mu \bigg]
\enqa
where   $S^c(x)$  and   $S^q(x)$  are   the  charm-   and  light-quark
propagators,  respectively.   The  next  step  is  to  write   the
correlation function in terms of a dispersion relation, such that
\begin{equation}
\Pi^{\rsup{OPE}}(q^{2}) = \int\limits_{4m^{2}_c}^{\infty}
 \!ds \:\frac{\rho^{\rsup{OPE}}(s)}{s - q^{2}},
\end{equation}
where $\rho^{\rsup{OPE}}(s)$  is given  by the  imaginary part  of the
correlation   function:  $\pi   \:\rho^{\rsup{OPE}}(s)  =   \mbox{Im}[
  \:\Pi^{\rsup{OPE}}(q^2 =  s) \:]$.  According  to Eq.~(\ref{fcope}),
the expression for the spectral density is 
\begin{eqnarray}
  \rho^{\rsup{OPE}}(s) &=& \!\frac{1}{2} 
  \qq^2 \mbox{cos}^2\theta ~\rho_{_{\chi_{c0}}}(s) + 
  \mbox{sin}^2\theta ~\rho_{_{D^\ast \! D^\ast}}(s) \nn\\
  && -~ \frac{\qq}{\sqrt{2}} \:\mbox{sin}\,\theta \:\mbox{cos}\,\theta ~
  \rho_{_{mix}}(s) ~~.
\end{eqnarray}
One calculates  the sum rule at  leading order in $\alpha_{s}$  in the
operators and considers  the contributions from the  condensates up to
dimension-8 in the  OPE. The expressions for the  spectral density are
given in Appendix \ref{apa}.

To improve  the matching between  the two sides  of the sum  rule, one
performs  the  Borel  transform.   After  transferring  the  continuum
contributions  to  the   OPE  side,  the  sum  rule   for  the  scalar
charmonium-molecule,     considered     as      a     mixed     scalar
$(\chi_{c0})\!-\!(D^\ast \!\bar{D}^\ast)$ state, can be written as
\begin{align}
\lambda_Y^{2} \:e^{- {M_Y^2}/M_B^{2}} &= 
\int\limits^{s_0}_{4m_{c}^{2}} \!\!ds \:e^{-s/M_B^2} ~\rho^{\rsup{OPE}}(s) ~~.
\label{sumrule}
\end{align}
Therefore, one can  estimate the ground state mass  from the following
ratio
\begin{equation}
{\cal R} = \frac{\int\limits_{4m_c^2}^{s_0} \!\! ds \:s \:e^{-s/M_B^2} ~\rho^{\rsup{OPE}}(s)}
{\int\limits_{4m_c^2}^{s_0} \!\! ds \:e^{-s/M_B^2} ~\rho^{\rsup{OPE}}(s)}
\label{ratio}
\end{equation}
where at the $M_B^2$-stability point, one obtains
\begin{eqnarray}
  M_Y &\simeq& \sqrt{{\cal R}} ~~.
\end{eqnarray}

\subsection{Numerical Analysis}

The numerical values  for the quark masses and  condensates are listed
in Table \ref{Param}.  These values are consistent with  the ones used
in  Refs.~\cite{x3872rad,x3872prod,diasY4260} for  the QCDSR  analysis on
other mixed hadronic states.

\begin{table}[h]
\setlength{\tabcolsep}{1.25pc}
\caption{\small QCD input parameters.}
\begin{tabular}{ll}
&\\
\hline
Parameters&Values\\
\hline
%
$\overline{m}_c$ & $(1.23 - 1.47) \GeV$ \\
$\qq$ & $ \hspace{-0.25cm}-(0.23 \pm 0.03)^3\GeV^3$\\
$\gGG$ & $(0.88 \pm 0.25)~\GeV^4$\\
$\gGGG$ & $(0.58 \pm 0.18)~\GeV^6$\\
$m_0^2 \equiv \qGq/\qq$ & $(0.8 \pm 0.1) ~\GeV^2$\\
$\rho \equiv \qqqq/\qq^2$ & $(0.5 - 2.0)$\\
\hline
\end{tabular}
\label{Param}
\end{table}

For reliable results  in a sum rule calculation, one  must establish a
valid Borel  window which  guarantees the existence  of a  region with
$M_B^2$-stability,  a good  OPE  convergence and  pole dominance  over
continuum contributions.  Nevertheless, another  crucial point  is the
optimal choice of  the continuum threshold $s_0$ and  the mixing angle
$\theta$.

%
\begin{figure}[t] 
\begin{center}
\centerline{\includegraphics[width=0.48\textwidth]{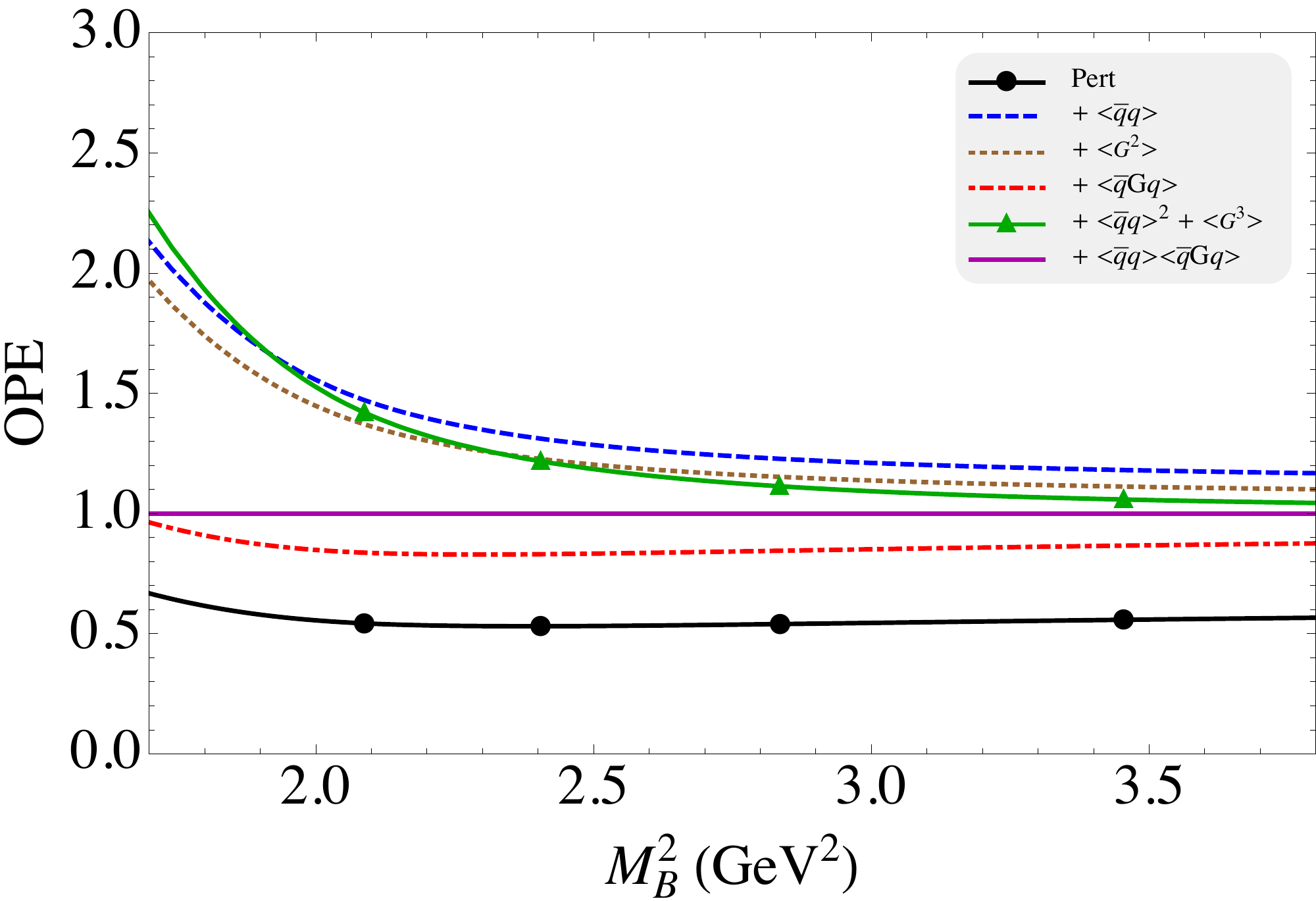}}
\caption{OPE convergence  in  the   region  $1.7 \leq   M_B^2  \leq
  3.8~\GeV^{2}$   for  $\sqrt{s_0}   =  4.40   \GeV$  and   $\theta  =
  76.0\degree$.  One plots  the  relative contributions  starting with  the
  perturbative contribution  (line with circles), and  each other line
  represents  the  relative contribution  after  adding  of one  extra
  condensate in the expansion: +  $\qq$ (dashed line), + $\GG$ (dotted
  line), +  $\qGq$ (dot-dashed  line), + $\qq^2$  + $\GGG$  (line with
  triangles) and $\qq \cdot \qGq$ (solid line).}
\label{fig1} 
\end{center}
\end{figure}  
\begin{figure}[t]
\begin{center}
\vspace{0.2cm}
\centerline{\includegraphics[width=0.48\textwidth]{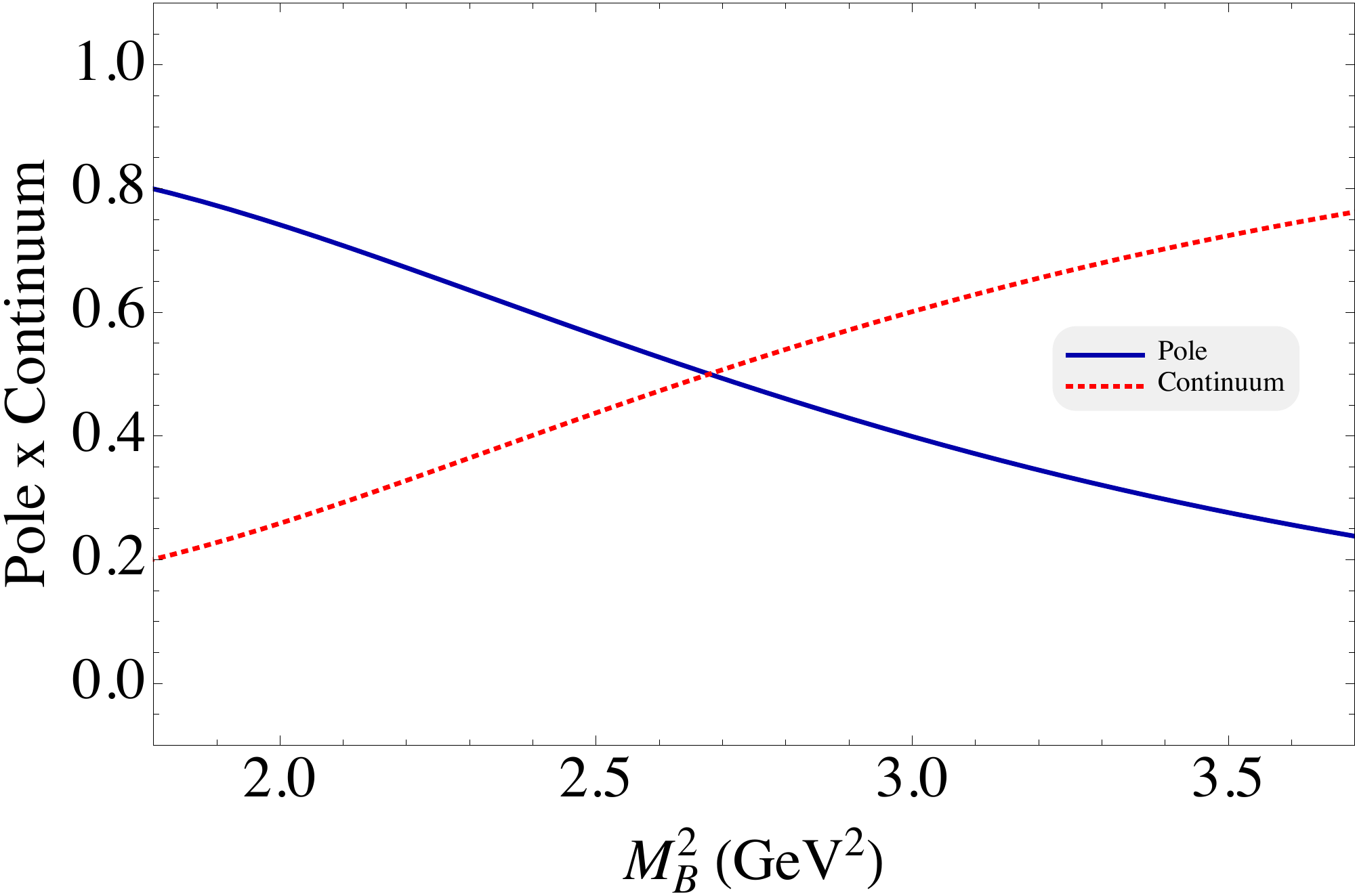}}
\caption{The   pole  (solid   line)   and   continuum  (dotted   line)
  contributions for $\sqrt{s_0} = 4.40 \GeV$ and $\theta = 76.0\degree$.}
\label{fig2} 
\end{center}
\end{figure}  

We  start  our  analysis  discussing   the  possible  values  of  both
parameters.  Considering that we are  interested in a mixed state with
a mass $M_Y \!\sim\! 3.9 \GeV$,  a reasonable initial value for the
continuum  threshold  would be  $\sqrt{s_0}  = 4.40  \GeV$.   In
principle,   the   choice   of   the  mixing   angle   seems   to   be
arbitrary. Hence,  for a fixed  value of  $\theta$, we search  for a
continuum   threshold  which   allows   us  to   determine  the   best
$M_B^2$-stability  inside  of a  valid  Borel  window.  After  lengthy
numerical calculations, we find  that the optimal choice is
\begin{eqnarray}
  \sqrt{s_0} &=& (4.40 \pm 0.10) \GeV \\
  \theta &=& (76.0 \pm 5.0)\degree ~~.
\end{eqnarray}
We notice that the OPE does not converge for $\theta$ values outside this
range.
Using these values, we analyze the relative contributions of the terms
in the OPE,  for $\sqrt{s_0} = 4.40 \GeV$ and  $\theta=76.0\degree$. As one
can  see in  Fig.~\ref{fig1},  the contribution  of the  dimension-8
condensate is smaller than $20$\% of the total contribution for values
of $M_B^2 \geq  2.4 \GeV^2$, which indicates the starting  point for a
good OPE  convergence. 
In order  to determine the  maximum value of  the Borel mass parameter,  we must
analyze  the  pole contribution.  Since  the  QCDSR approach  extracts
information only from the ground state,  we have to ensure that the pole
contribution is greater than the continuum contribution. Thus, we fix
the maximum value of the Borel mass parameter as the value for which the pole is
greater than or equal to the continuum contribution.
From  Fig.~\ref{fig2}, we can see that this condition is satisfied when  
$M_B^{2}  =  2.7 \GeV^2$. Therefore, the  Borel window is set as 
$2.4  \leq M^{2}_{B} \leq 2.7$ GeV$^{2}$.

\begin{figure}[t]
\begin{center}
\centerline{\includegraphics[width=0.48\textwidth]{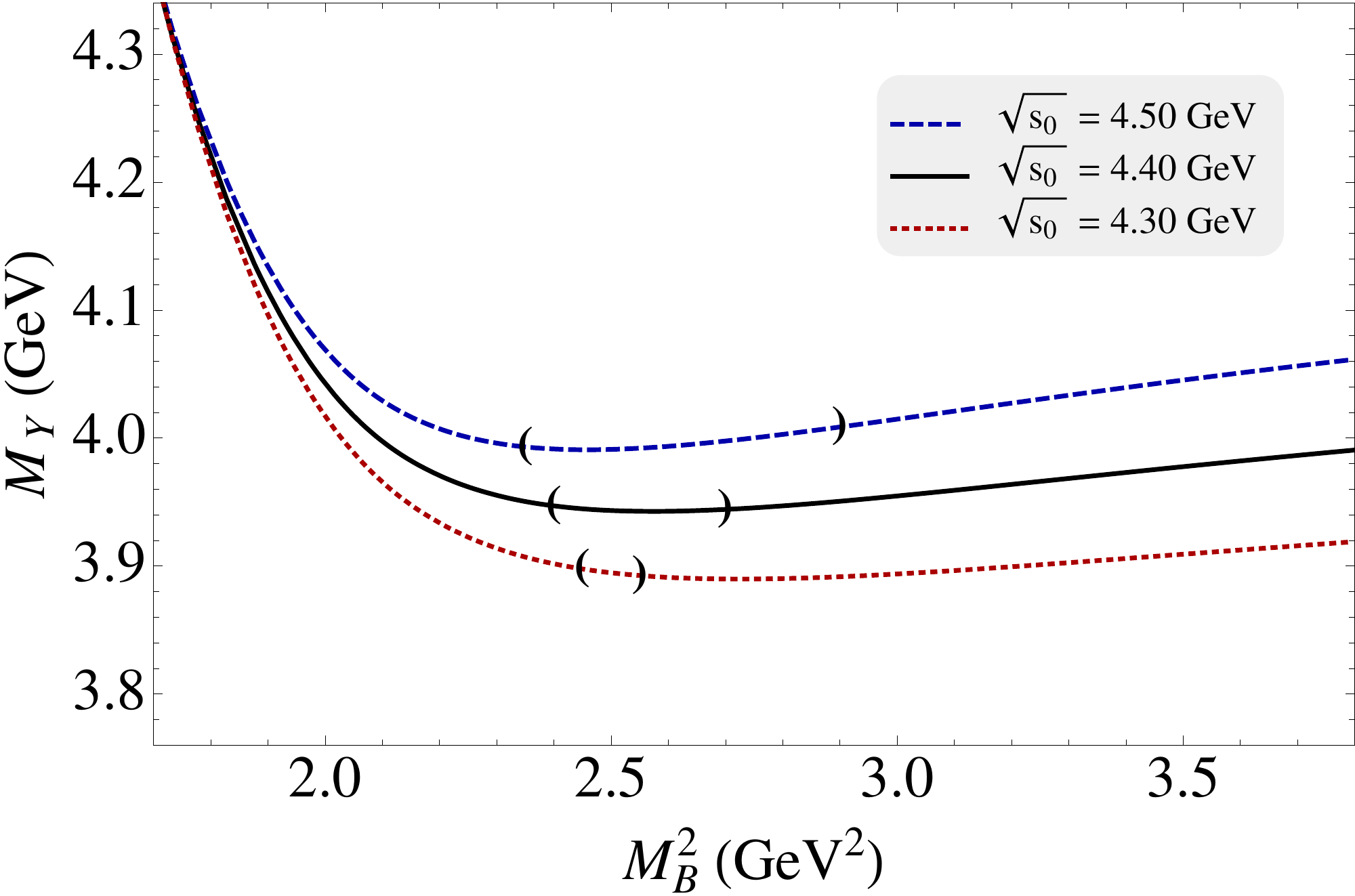}}
\caption{The mass as a function of  the sum rule parameter $M_B^2$ for
  $\sqrt{s_0} =  4.30 \GeV$  (dotted line),  $\sqrt{s_0} =  4.40 \GeV$
  (solid  line)  and  $\sqrt{s_0}  = 4.50  \GeV$  (dashed  line).  The
  respective parentheses indicate the valid Borel window.}
\label{fig3} 
\end{center}
\end{figure}

In Fig.  \ref{fig3}, we plot  the ground state  mass as a  function of
$M_B^2$,  considering  three  different  values  of  $\sqrt{s_0}$.  We
conclude that there is a good $M_B^2$-stability in the determined Borel
window.

Varying  the   value  of   the  continuum   threshold  in   the  range
$\sqrt{s_{0}} = (4.40  \pm 0.10) \GeV$, the mixing angle  in the range
$\theta=(76.0\pm  5.0)\degree$, and  the other  parameters as  indicated in
Table \ref{Param}, we get
\begin{equation}
M_{Y} = (3.95 \pm 0.11) ~ \mbox{GeV} ~.
\end{equation}
This mass  is compatible with  the experimental mass of  the $Y(3940)$
state  observed by  {\sc Belle} Collaboration  \cite{belleY3940}. Therefore,
from   a   QCD   sum   rule    point   of   view,   a   mixed   scalar
$(\chi_{c0})\!-\!(D^\ast  \!\bar{D}^\ast)$  state   could  be  a  good
candidate to explain the $Y(3940)$ state.

After  the determination  of  the  mass, we  can  use  this result  in
Eq.~(\ref{sumrule})  to estimate  the coupling  parameter, defined  in
Eq.~(\ref{coupling}). Therefore, considering the same values of $s_0$,
$\theta$ and  the Borel window  used for the  mass calculation,  we obtain
\beq \lambda_Y = (2.1 \pm 0.6) \times 10^{-2} ~ \mbox{GeV}^5.
\label{lay}
\enq

%

\section{ The $Y(3940) \to J/\psi \:\omega$ decay width}

In order  to provide more evidence  to support the conclusion  reached at
the end of  the previous section, that the $Y(3940)$  can be explained
as  a scalar mixed  state, we  now  use the  QCDSR to  compute the form factor
associated  with the vertex $Y\:J/\psi\:\omega$  and to estimate
the width  of the channel $Y(3940) \to  J/\psi \:\omega$.   For this
purpose, we start writing  the three-point function defined as
\beq \Pi_{\mu \nu}(p,\pli,  q) = \int
d^4x \:d^4y ~e^{i\pli \cdot x} \:e^{iq\cdot y} ~\Pi_{\mu \nu}(x,y),
\label{3po}
\enq
where $p=\pli+q$ and $\Pi_{\mu \nu}(x,y)$ is given by
\beq
\Pi_{\mu \nu}(x,y)=\lag 0|T\{j_{\mu}^{\psi}(x)
j_{\nu}^{\omega}(y)j^{\dagger}(0)\}|0\rag.
\label{pixy}
\enq
The  interpolating  currents for  the  $J/\psi$  meson and  the  mixed
$(\chi_{c0})\!-\!(D^\ast     \!\bar{D}^\ast)$     state    used     in
Eq.~(\ref{pixy}) are defined in  Section \ref{mixoperator},
while the interpolating current associated  with the $\omega$ meson is
defined by
\beq
j^{\omega}_\nu = \frac{1}{6}\Big(\bar{u}_a \gamma_\nu u_a +
\bar{d}_a \gamma_\nu d_a \Big).
\label{jomega}
\enq 

In the  same manner that it was done for the two-point correlation
function,  we  again invoke  the quark-hadron duality principle to  calculate 
the three-point  function in two ways. We match both  sides
after performing the Borel  transform.  In the phenomenological  side, one has
to  insert the  intermediate  states for the $J/\psi$, $\omega$  and
$Y(3940)$ mesons in Eq.~(\ref{3po}).  Using the following relations:
\beqa
\lag 0 |\: j_\mu^\psi \:| J/\psi(\pli)\rag &=& M_\psi 
f_{\psi} \:\epsilon_\mu(\pli), \nn\\
\lag 0 |\: j^\omega_\nu \:| \omega(q)\rag &=& M_\omega 
f_{\omega} \:\epsilon_\nu(q), \\
\lag Y(p) |\: j \:| 0\rag &=& \la_Y, \nn
\enqa
we obtain the expression
\beqa
\Pi_{\mu\nu}^{\rsup{PHEN}} (p,\pli,q)&=&{\la_Y \:M_{\psi}
f_{\psi} \:M_\omega f_\omega ~ g_{_{\!Y \!\psi \omega}}(q^2)
\over (p^2-M_{Y}^2)({\pli}^2- M_{\psi}^2)(q^2- M_\omega^2)} ~~ \nn \\
&&\times ~\Big[ q_{\mu} p^\prime_\nu 
-(\pli \cdot q)g_{\mu\nu} \Big] +\cdots\;,
\lb{phen}
\enqa
where  the dots  stand for  the contribution  of all  possible excited
states.  The form factor, $g_{_{\!Y \!\psi \omega}}(q^2)$, is defined by
the generalization of  the on-shell mass matrix  element, $\lag J/\psi
\:\omega \:|\: Y \rag$, for an off-shell $\omega$ meson:
\beqa
\lag J/\psi \:\omega \:|\: Y\rag &=&g_{_{\!Y \!\psi \omega}}(q^2) \bigg[
\Big( \pli \cdot \epsilon^*(q) \Big)   \Big(q \cdot \epsilon^*(\pli) \Big) \nn \\
&& -~ (\pli \cdot q) \Big(\epsilon^*(\pli) \cdot \epsilon^*(q) \Big) \bigg],
\label{coup}
\enqa
which can  be extracted from  the effective Lagrangian  that describes
the coupling between two vector mesons and one scalar meson:
\beq
{\cal{L}}=\frac{i}{2} \:g_{_{\!Y \!\psi \omega}} V_{\alpha\beta} \Psi^{\alpha \beta}~Y
\enq
where   $V_{\alpha  \beta}   =   \partial_{\alpha}  \omega_{\beta}   -
\partial_{\beta}   \omega_{\alpha}$   and   $\Psi^{\alpha   \beta}   =
\partial^{\alpha} \psi^{\beta} -  \partial^{\beta} \psi^{\alpha}$, are
the tensor fields of the $\omega$ and $\psi$ fields respectively.

In  the OPE  side, we  calculate the  correlation function  at leading
order in  $\alpha_s$ and  we consider condensates  up to  dimension 7.
Notice that the three-point function includes a number of different Lorentz 
structures and the most suitable one for our purposes seems to be 
the $q_{\mu}p^\prime_\nu$. The reasons for the choice of this structure are: 
(a) it has the larger number of momenta; (b) the OPE leading 
term decreases as $1/Q^2$ as $Q^2 \rightarrow \infty$, which is an expected 
behavior for QCD form factors. 
In general, for any given structure, the sum rule method is inapplicable at 
large $Q^2$ where the power corrections become large and uncontrollable.
At small $Q^2$, the situation is even worse since when approaching the physical 
region the operator expansion stops working. 
In this sense, one has to consider that the sum rule is valid up to a rather 
small $Q^2$ and the extrapolation from the values of $Q^2$ to the physical 
region can be obtained with a good accuracy.

Matching both side of the sum rule, taking the approximation 
$p^2  \simeq {\pli}^2=-P^2$ and doing the Borel transform to 
$P^2 \rightarrow M_B^2$,  we get the following expression in the 
$q_\mu p^\prime_{\nu}$ structure:
\beqa
\frac{\lambda_Y M_{\omega} f_{\omega} M_\psi f_\psi ~g_{_{\!Y\!\psi \omega}}(Q^2)}
{(M_Y^2 - M_\psi^2)(Q^2+M_\omega^2)} \!\!\!\!\!\!\!&& \:\left(e^{-M_Y^2/M_B^2}
-e^{-M_\psi^2/M_B^2}\right)+\nn\\
+H(Q^2)~e^{-s_0/M_B^2} &=& \Pi^{\rsup{OPE}}(M_B^2,Q^2), 
\label{3sr}
\enqa
where  $Q^2=-q^2$,   and  $H(Q^2)$ function  represents  the   contribution  to  the
pole-continuum       transitions      \cite{matheusX3872,io2,decayx,dsdpi}.
The $\Pi^{\rsup{OPE}}(M_B^2,Q^2)$ function is
\beqa
\Pi^{{\rsup{OPE}}}(M_B^2,Q^2)=\sin\theta \int\limits_{4m_c^2}^{+\infty} \!\!ds ~
e^{-s/M_B^2} \:\rho(s,Q^2) ~,
\label{opeside}
\enqa
and $\rho = \rho^{pert} + \rho^{\qq} + \rho^{\GG} + \rho^{\qGq} + \rho^{\qq \GG}$ 
is given explicitly by 
\begin{eqnarray}
  \rho^{pert}(s,Q^2) &=& ~~\rho^{\qq}(s,Q^2) ~~=~~ 0 ~, \\ \nn\\
  \rho^{\GG}(s,Q^2) &=&  -\frac{\gGG}{3^2 \cdot 2^{10} \:\pi^4} 
  \int\limits_0^1 \!d\al   ~\delta \!\left[ s - \frac{m_c^2}{\alpha (1 \!-\! \alpha )} \right] \nn\\
  && \times~(3-3\alpha +\alpha^2) ~,\\ \nn\\
  \rho^{\qGq}(s,Q^2) &=&  \frac{m_c \qGq}{72 \pi^2 Q^2}
  \int\limits_0^1 \!d\al   ~\delta \!\left[ s - \frac{m_c^2}{\alpha (1 \!-\! \alpha )} \right] ~,\nn\\
  \rho^{\qq\GG}(s,Q^2) &=& \frac{m_c \qq \gGG}{3^3 \cdot 2^5 \pi^2 Q^4} 
   \int\limits_0^1 \!d\al   ~\delta \!\left[ s - \frac{m_c^2}{\alpha (1 \!-\! \alpha )} \right] \nn\\
  &&\times~ \frac{(1- 3\al + 3\alpha^2)}{\alpha (1-\alpha )} ~~.
\end{eqnarray}
As observed in previous works 
\cite{matheusX3872,x3872rad,x3872prod,diasY4260}, the charmonium part
of  the  mixed  current  defined in  Eq.~(\ref{mixj})  contributes  to the
three-point  function uniquely with disconnected diagram. Hence, only the
molecule part contributes to the decay channel $Y(3940)\to J/\psi\: \omega$.
This fact is evident due to the presence of the sine function in Eq.~(\ref{opeside}).

We follow the usual procedure in order to extract  the value of the
coupling constant  associated with the $Y\to J/\psi \:\omega$ process. 
First,  we must determine the  form factor of the $Y\:J/\psi\:  \omega$
vertex, which  can  be  done  by  isolating  the  function
$g_{_{\!Y \!\psi\omega}}(Q^2)$   in   Eq.~(\ref{3sr}),    then we  divide
Eq.~(\ref{3sr}) by its derivative with respect to  $1/M_B^2$ in order
to eliminate the  unknown function $H(Q^2)$.  Therefore, we are left with a 
function for the form factor
$g_{_{\!Y\!\psi \omega}}(Q^2)$ to be determined numerically.

In the  numerical analysis we use the experimental values of the
meson masses and decay constants: $M_{\psi}=3.10 \GeV$, $f_{\psi}=0.405$
GeV, $M_{\omega}=0.782$  GeV, $f_{\omega}=0.046$ GeV. For the Y mass,
we  use  the  experimental  value in  Ref.~\cite{belleY3940}  and  the
meson-current parameter  $\lambda_Y$ which has been evaluated in 
the previous section, see Eq.~(\ref{lay}).

\begin{figure}[t]
\begin{center}
\centerline{\includegraphics[width=0.5\textwidth]{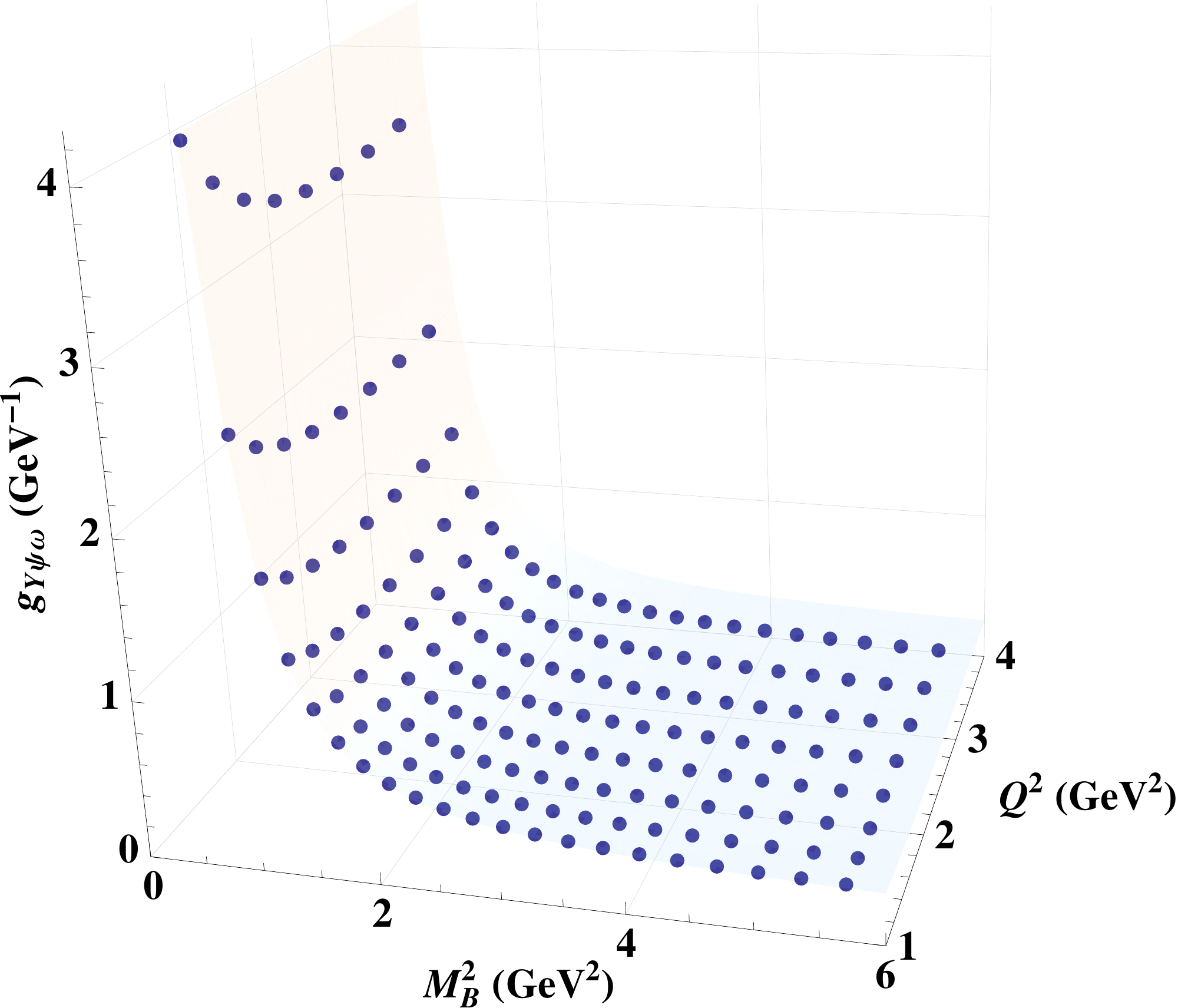}}
\caption{The form factor $g_{_{\!Y \!\psi\omega}}(Q^2)$ as a function of 
the momentum $Q^2$ and Borel mass parameter $M_B^2$.}
\label{fig3D} 
\end{center}
\end{figure}
\begin{figure}[t]
\begin{center}
\centerline{\includegraphics[width=0.45\textwidth]{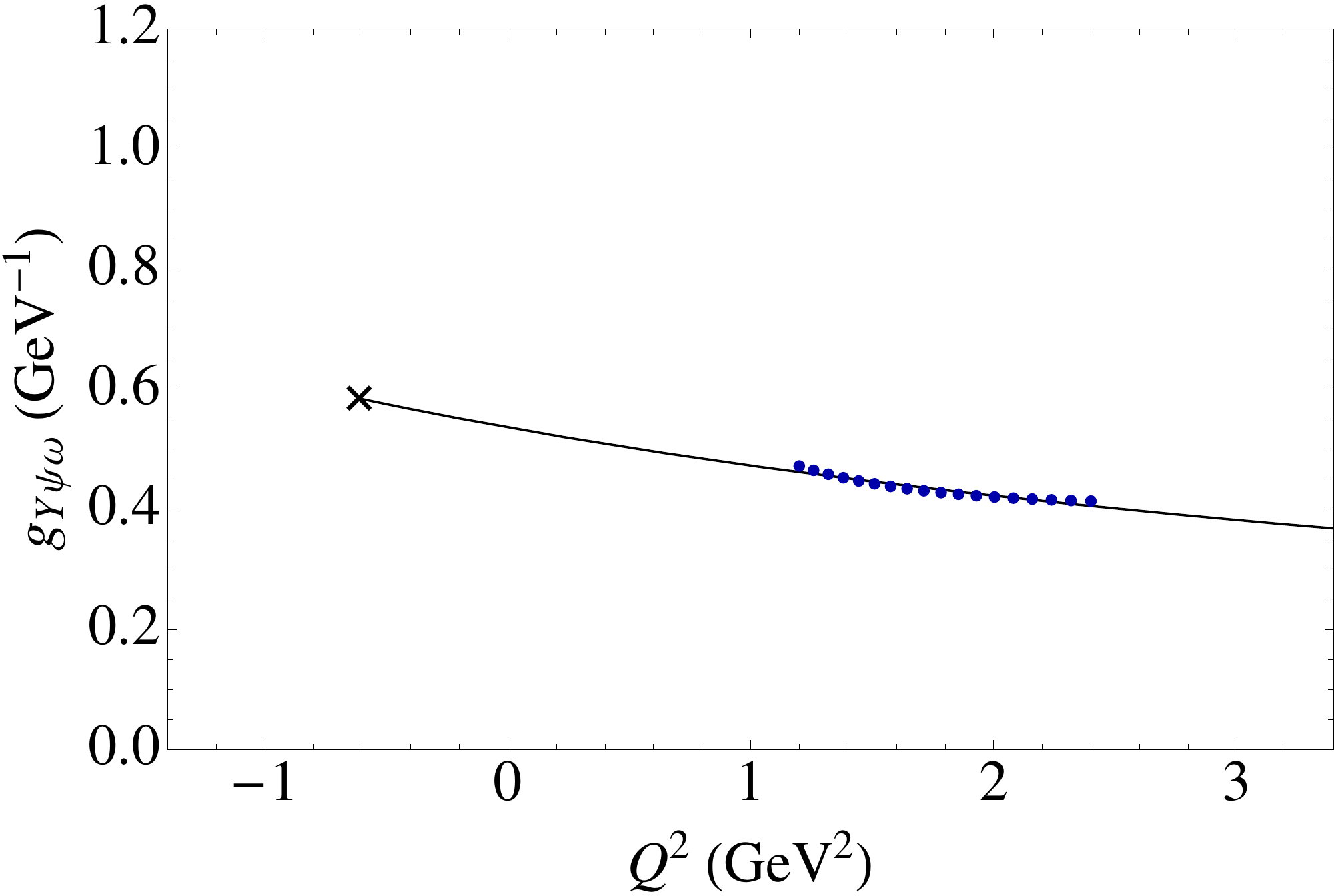}}
\caption{QCDSR  results  for the form factor  $g_{_{\!Y \!\psi\omega}}(Q^2)$, 
  for  $\sqrt{s_0}=4.40$ GeV  (circles).  The solid  line gives
  the    parametrization     of    the    QCDSR     results    through
  Eq.~(\ref{expform}).  The  cross  indicates   the  value  of  the  coupling
  constant.}
\label{fig2D} 
\end{center}
\end{figure} 

In   Fig.   \ref{fig3D},  we  show   a   plot  of   the  form   factor
$g_{_{\!Y \!\psi\omega}}(Q^2)$ as a function of $M_B^2$ and $Q^2$.  
Note that, a reliable sum rule  must be independent of the choice of Borel 
mass parameter.
As one can see, we obtain a good stability in Borel mass parameter 
at $M_B^2\geq 1.8 \GeV^2$. Here we work at the interval a 
$1.8~\GeV^2\leq M_B^2\leq 4.0~\GeV^2$.
The form factor dependence in $Q^2$ can be evaluated by taking the 
average of the $M_B^2$ values inside this stability region. The results are 
shown in Fig.~\ref{fig2D}.

As mentioned above,  the sum rule is not reliable at very large and 
very small values of $Q^2$. Here we find that the 
results are reliable for $1.2 \leq Q^2 \leq 2.4 \GeV^2$.

Once we have  determined the form factor behavior, we can now extract the coupling
constant by using the momentum value at the   omega    meson   pole,
$Q^2=-M_\omega^2$. For this purpose, we  have to extrapolate the form
factor  to the  region of  $Q^2$ where  the QCDSR  is not  valid.  This
extrapolation can be done by  parametrizing the QCDSR results shown in
Fig.~\ref{fig2D} for $g_{_{\!Y \!\psi\omega}}(Q^2)$ using a monopolar function:
\beq
g_{_{\!Y \!\psi \omega}}(Q^2) = \frac{g_1}{g_2 + Q^2},
\label{expform}
\enq
and the results for the fitting parameters are:
\beqa
&&g_1 = (4.0~\pm ~1.0)~\mbox{GeV};\nonumber\\ 
&& g_2=(7.4 ~\pm ~ 0.2)~\mbox{GeV}^2.  
\label{c1c2}
\enqa

The theoretical errors are evaluated considering errors on the following
parameters: $\sqrt{s_0}=4.40\pm0.10$  GeV, $\theta=76.0^\circ\pm5.0^\circ$,
and also the error on  the meson coupling parameter $\lambda_Y$, given
by Eq.~(\ref{lay}).  We notice that the results do not depend much on the
parameters $\sqrt{s_0}$ and $\theta$, while the theoretical errors are
mainly affected by the meson coupling $\lambda_Y$.

In order to see how well  the parametrization works, the solid line in
Fig.~\ref{fig2D} represents the  Eq.~(\ref{expform}) with values given
by Eq.~(\ref{c1c2}).
The   coupling  constant,   $g_{_{\!Y \!\psi\omega}}$ ,  is   given  by   using
the momentum value $Q^2=-M^2_{\omega}$ in Eq.~(\ref{expform}). 
Then, we get:
\beq
g_{_{\!Y \!\psi\omega}}=g_{_{\!Y \!\psi\omega}}(-M^2_\omega)=
(0.58~ \pm~0.14 )~~\mbox{GeV}^{-1}.
\label{coupvalue}
\enq

The decay width for this process $Y(3940) \rightarrow J/\psi\: \omega$ is
given by
\beqa
\Gamma_{{Y(3940) \to J/\psi\:\omega}}= \frac{g^2_{_{\!Y \!\psi \omega}}}{3}
{p(M_Y,M_\omega ,M_\psi)\over8\pi M_{Y}^2} \hspace{1cm}
\nn\\
\times \left(M_\psi^2 M_\omega^2 + 
{1\over2}(M_Y^2- M^2_{\psi}-M_\omega^2)^2\right),\label{gamma}
\enqa
where
\beq
p(a,b,c) \equiv {\sqrt{a^4+b^4+c^4-2a^2b^2-2a^2c^2-2b^2c^2}\over 2a}.
\enq

Therefore,  we  obtain the  decay width inserting the  value obtained  for  the
coupling constant (\ref{coupvalue}) in (\ref{gamma}):
\beq
\Gamma_{Y(3940)\to J/\psi \:\omega}=(1.7~\pm ~0.6)~\MeV.
\label{width}
\enq
This result is consistent with the experimental width of the state and
the   lower    limit   for    the   process    $Y\to   J/\psi\:\omega$
\cite{belleY3940,babarY3940,godfrey,godfrey2}.  It is also of the same
order as other available theoretical evaluations \cite{oset,gutsche}.

\section{Summary and Conclusions}

In summary, we have used the QCDSR approach to study the two-point and
three-point functions of  the $Y(3940)$ state, by  considering it as a mixed
charmonium-molecule state.  We have  evaluated the mass working with
the  two-point  function  at  leading order  in  $\alpha_{s}$  and  we
consider the contributions from  the condensates up to dimension-8.
 We obtained a mass which is in a very good agreement with the experimental 
value for the $Y(3940)$ state, and we found a mixing angle around 
$\theta = (76.0\pm5.0)^{0}$.

To evaluate  the width  of the  decay channel $Y(3940)\to  J/\psi\:\omega$, we
work  with  the   three-point  function  also  at   leading  order  in
$\alpha_{s}$ and we consider the contributions from the condensates up
to  dimension-7.   The obtained  value  of the width  is  $\Ga_{Y\to
  J/\psi \:\omega} = (1.7\pm 0.6)$ MeV,  which is  smaller than
the total  experimental width \cite{belleY3940,babarY3940}, but  is
consistent     with     the     lower     limit for this channel     
$\Gamma>1$     MeV
\cite{oset,gutsche}.   Thus, according to the available experimental data, 
we can conclude that a mixing between the $\chi_{c0}$ charmonium and 
the $D^\ast \!\bar{D}^\ast$ molecule, could be a good candidate to explain the 
$Y(3940)$ state.

\subsection*{Acknowledgment} 
This work has been supported by FAPESP and CNPq.

\appendix

\section{Spectral Densities for the Two-point Correlation Function}
\label{apa}

Next,  we   list  the   spectral  densities   for  the   mixed  scalar
$(\chi_{c0})\!-\!(D^\ast  \!\bar{D}^\ast)$  state   described  by  the
current in Eq.~(\ref{mixj}).  We consider the OPE  contributions up to
dimension-8 condensates and keep terms at leading order in $\alpha_s$.
In  order  to  retain  the  heavy   quark  mass  finite,  we  use  the
momentum-space  expression   for  the  heavy  quark   propagator.   We
calculate  the light  quark part  of the  correlation function  in the
coordinate-space and use the Schwinger parametrization to evaluate the
heavy  quark part  of the  correlator. For  the $d^4x$  integration in
Eq. (\ref{fc}),  we use again  the Schwinger parametrization,  after a
Wick rotation.  Finally, the  result of these  integrals are  given in
terms of logarithmic  functions through which we  extract the spectral
densities. The  same technique can be used for evaluating the 
condensate contributions.

For  the $\chi_{c0}$  meson contribution,  the spectral  densities are
written below \cite{rry}
\begin{eqnarray*}
  \rho^{pert}_{_{\chi_{c0}}}(s) &=& -\frac{3 m_c^2}{8\pi^2} ~
  v \left(4-\frac{1}{x}\right) ~~, 
\end{eqnarray*}
\begin{eqnarray*}
  \rho^{\GG}_{_{\chi_{c0}}}(s) &=& \frac{\gGG}{2^4 \pi^2} 
  \int\limits^{1}_{0} \!d\al \:\de \!\left[ s- \frac{m_c^2}{\al(1-\al)} \right] \nn\\
  && \times~ \bigg( 1 - \frac{m_c^2/M_B^2}{\al(1-\al)^2} \bigg) ~,\\ \\ 
  \rho^{\GGG}_{_{\chi_{c0}}}(s) &=& \frac{m_c^2 \:\gGGG}{3 \cdot 2^6 \pi^2 \:M_B^4} 
  \int\limits^{1}_{0} \!d\al \:\de \!\left[ s- \frac{m_c^2}{\al(1-\al)} \right] \nn\\
  && \times~ \bigg(\frac{72-83\al+14\al^2}{\al(1-\al)^3} \bigg) ~~. \\ \\ 
\end{eqnarray*}
For the $D^\ast \bar{D}^\ast ~(0^{++})$ molecular state \cite{albuquerque}
\begin{eqnarray*}
  \rho^{pert}_{_{D^\ast \! D^\ast}}(s) &=& \frac{m_c^8}{5\cdot 2^{12} \pi^6}
  \bigg[ v \!\left( 480 + \frac{1460}{x} - \frac{274}{x^2} - \frac{38}{x^3} + \frac{1}{x^4} \right) \\
  && \hspace{-1.9cm} ~+ 120\mathcal{L}_v \left( 8x - 1 - 
  6\mbox{\:Log}(x) - \frac{8}{x} + \frac{2}{x^2} \right) - 1440\mathcal{L}_+ \bigg] ~,\\ \\
  \rho^{\qq}_{_{D^\ast \! D^\ast}}(s) &=& \frac{m_c^5 \qq}{64 \pi^4} 
  \bigg[ v \!\left( \!6 \!-\! \frac{5}{x} \!-\! \frac{1}{x^2} \!\right) \!+\! 
  6\mathcal{L}_v \!\left( \!2x \!-\! 2 \!+\! \frac{1}{x} \!\right) \!\bigg] ~, \\ \\
  \rho^{\GG}_{_{D^\ast \! D^\ast}}(s) &=& \frac{m_c^4\gGG}{3 \cdot 2^{10} \pi^6} 
  \bigg[ v \!\left( \!6 \!-\! \frac{5}{x} \!-\! \frac{1}{x^2} \!\right) \!+\! 
  6\mathcal{L}_v \!\left( \!2x \!-\! 2 \!+\! \frac{1}{x} \!\right) \!\bigg] ,  \\ \\ 
  \rho^{\qGq}_{_{D^\ast \! D^\ast}}(s) &=& \frac{3 m_c^3 \qGq}{128\pi^4} 
  \left( \frac{v}{x} - 2\mathcal{L}_v \right) ~,\\ \\ 
  \rho^{\qq^2}_{_{D^\ast \! D^\ast}}(s) &=& \frac{m_c^2 \:\rho \qq^2} {4\pi^2} ~v ~,\\ \\ 
  \rho^{\GGG}_{_{D^\ast \! D^\ast}}(s) &=& \frac{m_c^2 \gGGG}{3\cdot 2^{12}\pi^6} 
  \!\bigg[ v \!\left( \!6 \!-\! \frac{25}{x} \!+\! \frac{1}{x^2} \!\right) \!\!+\! 
  6\mathcal{L}_v \!\left( \!2x \!+\! 2 \!+\! \frac{1}{x} \!\right) \!\!\bigg] ,\\ \\ 
  \rho^{\langle 8 \rangle}_{_{D^\ast \! D^\ast}}(s)\!\! &=& -\frac{m_c^2 \qq \qGq}{8 \pi^2}
  \int\limits_0^1 \!d\al   ~\delta \!\left[ s - \frac{m_c^2}{\alpha (1 - \alpha )} \right] \\
  && \times~
  \left( 1 + \frac{m_c^2/M_B^2}{\alpha(1 - \alpha )} \right) ~~.\\
\end{eqnarray*}
Finally, for the mixed term, we have
\beqa
  \rho^{\qq}_{mix}(s) &=& \frac{m_c^2 \qq}{4 \pi^2} ~v \left( 4 - \frac{1}{x} \right) ~, \nn\\  
  \rho^{\qGq}_{mix}(s) &=& 0 ~~. \nn
\enqa
In all these expressions we have used the following definitions:
\begin{eqnarray}
  x &=& m_c^2/s ~~,\\ \nn\\
  v &=& \sqrt{1 - 4x} ~~,\\ \nn\\
  {\cal L}_v &=& \mbox{Log}\left( \frac{1+v}{1-v} \right) ~~,\\ \nn\\
  {\cal L}_+ &=& \mbox{Li}_2 \left( \frac{1+v}{2} \right) - \mbox{Li}_2 \left( \frac{1-v}{2} \right) ~~.
\end{eqnarray}
\vfill

\bibliographystyle{unsrt}

\end{document}